\documentclass[twocolumn,pra,preprintnumbers,amsmath,amssymb]{revtex4-1} 

\usepackage{ae,aecompl} 
 
\usepackage{graphicx} 
\usepackage[caption=false]{subfig}
\usepackage{multirow}
\usepackage{array}
\usepackage{amsfonts}

\usepackage[latin1]{inputenc} 
\usepackage[T1]{fontenc} 
 
\usepackage{color}  
 
\newcommand{\etal}{{\it et al.}}

\begin{document} 
 
\def\pss#1#2#3{Phys.~Stat.~Sol.~{\bf #1},\ #2\ (#3)} 
\def\apl#1#2#3{Appl.~Phys.~Lett.~{\bf #1},\ #2\ (#3)} 
\def\jpb#1#2#3{J.~Phys.~B:~{\bf #1},\ #2\ (#3)} 
\def\jpc#1#2#3{J.~Phys.~Chem.~{\bf #1},\ #2\ (#3)} 
\def\jpcb#1#2#3{J.~Phys.~Chem. B~{\bf #1},\ #2\ (#3)} 
\def\jcp#1#2#3{J.~Chem.~Phys.~{\bf #1},\ #2\ (#3)} 
\def\cpl#1#2#3{Chem.~Phys.~Lett.~{\bf #1},\ #2\ (#3)} 
\def\pr#1#2#3{Phys.~Rev~{\bf #1},\ #2\ (#3)} 
\def\pra#1#2#3{Phys.~Rev.~A~{\bf #1},\ #2\ (#3)} 
\def\prb#1#2#3{Phys.~Rev.~B~{\bf #1},\ #2\ (#3)} 
\def\prc#1#2#3{Phys.~Rev.~C~{\bf #1},\ #2\ (#3)} 
\def\prd#1#2#3{Phys.~Rev.~D~{\bf #1},\ #2\ (#3)} 
\def\pre#1#2#3{Phys.~Rev.~E~{\bf #1},\ #2\ (#3)} 
\def\rmp#1#2#3{Rev.~Mod.~Phys.~{\bf #1},\ #2\ (#3)} 
\def\prl#1#2#3{Phys.~Rev.~Lett.~{\bf #1},\ #2\ (#3)} 
\def\sci#1#2#3{Science~{\bf #1},\ #2\ (#3)} 
\def\nat#1#2#3{Nature~{\bf #1},\ #2\ (#3)} 
\def\apj#1#2#3{Astrophys.~J.~{\bf #1},\ #2\ (#3)} 
\def\njp#1#2#3{N.~J.~Phys.~{\bf #1},\ #2\ (#3)}
\def\phys#1#2#3{Physica~{\bf #1},\ #2\ (#3)}

\def\bea{\begin{eqnarray}} 
\def\eea{\end{eqnarray}} 
\def\be{\begin{equation}} 
\def\ee{\end{equation}} 
\def\etal{{\it et al.}} 
 
\newcommand{\JCP}[1]{J. Chem. Phys. {\bf #1}} 
\newcommand{\ApJ}[1]{Astrophys. J. {\bf#1}} 
\newcommand{\CPL}[1]{Chem. Phys. Lett. {\bf #1}} 
\newcommand{\PRA}[1]{Phys. Rev. A {\bf #1}} 
\newcommand{\JPC}[1]{J. Phys. Chem. {\bf #1}} 
\newcommand{\JPCA}[1]{J. Phys. Chem. A {\bf #1}} 
\newcommand{\PRL}[1]{Phys. Rev. Lett. {\bf #1}} 
\def\BE{\begin{equation}} \def\EE{\end{equation}} 
 
\preprint{Preprint} 
 
\title{Influence of  monolayer contamination on electric-field-noise heating in ion traps}  
\date{\today} 
\author{A. Safavi-Naini$^{1,2}$, E. Kim$^3$, P. F. Weck$^4$, P. Rabl$^5$, and H. R. Sadeghpour$^{2}$}
\affiliation{$^1$ Department of Physics, Massachusetts Institute of Technology, Cambridge, MA 02139}
\affiliation{$^2$ ITAMP, Harvard-Smithsonian Center for Astrophysics, Cambridge, Massachusetts 02138}
\affiliation{$^3$Department of Physics and Astronomy, University of Nevada, 
Las Vegas, NV 89154-4002} 
\affiliation{$^4$Sandia National Laboratories, P.O. Box 5800, Albuquerque, NM 87185-0779}
\affiliation{$^5$ Institute of Atomic and Subatomic Physics, TU Wien, Stadionallee 2, 1020 Wien, Austria  }


\begin{abstract} 
Electric field noise is a hinderance to the assembly of large scale quantum computers based on entangled trapped ions. 
Apart from ubiquitous technical noise sources, experimental studies of trapped ion heating have revealed additional limiting contributions to this noise, originating from atomic processes on the electrode surfaces. In a recent work [A. Safavi-Naini {\it et al.}, Phys. Rev. A {\bf 84}, 023412 (2011)]
we described a microscopic model for this excess electric field noise, which points a way towards a more  systematic understanding of surface adsorbates as progenitors of electric field jitter noise. Here, we address the impact of surface monolayer contamination on adsorbate induced noise processes. By using exact numerical calculations for H and N atomic monolayers on an Au(111) surface representing opposite extremes of physisorption and chemisorption, we show that an additional monolayer can significantly affect the noise power spectrum and either enhance or suppress the resulting heating rates.  

\end{abstract}
\date{\today} 
\maketitle 
 \section{Introduction}

Ion trap miniaturization and a precise control of errors in the entangled qubits are two key prerequisites for using trapped laser-cooled atomic ions as multi-qubit logic gates in a scalable quantum architecture ~\cite{HaeffnerPhysRep2008,WinelandLaserPhysics2011, Wineland1998}. One main source of qubit error in such systems is the motional jitter of the collective behavior of the ions in micro traps \cite{Turchette, DeVoe,IonTrapCooling2, IonHeating3,IonHeating2,MIT2008, MITsupercond11,DaniilidisNJP2011,Allcock,wineland2011}. Some of the unwanted heating noise is naturally mitigated by operating the traps at cryogenic temperatures \cite{IonHeating3,MIT2008,MITsupercond11}. However the noise still remains larger than the expected Johnson noise for the traps. Early observations of the dependance of the heating rate on position of the ions above the trap ($\sim 1/d^4$) \cite{Turchette} and on the elapsed time in the ion-loading region \cite{DeVoe, Turchette}, lend credence to the role played by surface contaminants. Further experiments with superconducting traps \cite{MITsupercond11} corroborate the understanding that the noise source lies on the surface and not in the bulk. The confirmation has come more directly from two recent and complimentary experiments, where laser cleaning \cite{Allcock} and ion beam bombardment \cite{wineland2011} of the trap electrodes led to a reduction of the noise. The experiment reported in \cite{wineland2011} also identified the surface contaminants as carbon based, in the form of 2-3 monolayers  (MLs) of hydrocarbons. 

Theoretical studies of the anomalous heating have been largely phenomenological, aiming to explain the signatures of this noise. These models use the concept of patch potentials developed by Turchette~\etal \ to explain the motional heating in ions \cite{Turchette, DaniilidisNJP2011,Dubessy2009,Chuang2011}. In a recent work \cite{Safavi-Naini2011} we developed a microscopic model to predict the features of electric field noise (distance, frequency and temperature dependencies) from the details of atomic surface processes. 
 This model is predicated on the idea that the noise in ion traps emanates from a random distribution of fluctuating dipoles associated with individual adatoms on a metallic electrode surface. 

 In this work we extend our earlier theoretical treatment by investigating the dependence of this surface noise 
on the presence of an additional ML of atomic species on a gold surface. To do so we present detailed numerical calculations on the adsorbate surface potential in the presence or absence of MLs with different reactivity.  
%
To this end, we chose He as the ML atom to represent weak binding (physisorbed species) and N as the ML atom for strong binding (chemisorbed species). For computational simplicity we choose hydrogen (H) as our adsorbate and compare the binding potential of 
${\rm H}/{\rm Au}(111)$, ${\rm H}/{\rm He}/{\rm Au}(111)$ 
and ${\rm H}/{\rm N}/{\rm Au}(111)$. Using density functional theory (DFT) we obtain accurate data for these surface potentials, induced dipole moments and modifications on the phonon density of states. Combined with the semi-analytical treatment of Ref.~\cite{Safavi-Naini2011} we use this data to extrapolate the resulting impact on phonon induced dipole fluctuations and find that MLs of different reactivity can lead to the completely opposite effects of reducing or enhancing the noise. 

The remainder of the paper is structured as follows. In Sec.~\ref{sec:Heating} we briefly review the problem of anomalous heating in ion traps and summarize the basic assumptions of the model detailed in Ref.~\cite{Safavi-Naini2011}. As the main part of this work we present in Sec.~\ref{methods} our numerical results on adatom surface potentials and induced dipole moments for two different types of MLs on Au. Finally, in Sec.~\ref{sec:noise} we discuss the impact of these findings on the adatom dipole fluctuation spectrum and conclude in Sec.~\ref{sec:conclusions}.

\section{Anomalous ion heating from fluctuating surface dipoles}\label{sec:Heating}

In micro-fabricated surface ion traps, which are currently developed for quantum information processing, single or multiple ions are trapped by electric potentials at a distance $d$ of a few $100\,\mu$m above a metal electrode. 
The resulting trapping frequencies are typically around $\omega_t \sim 1$ MHz and allow efficient laser cooling and coherent manipulations of the trapped ion. However, when cooled close to the quantum ground state of the trap, the ion motion is still associated with a comparably large electric dipole moment $d_I \approx  q a_0$, where $q$ is the charge and $a_0=\sqrt{\hbar/(2m_I \omega_t)}$ the zero-point motion for an ion mass $m_I$. 
Fluctuating electric fields from the environment couple to this dipole moment and excite the ion motion with a characteristic heating rate~\cite{Turchette}
 \begin{equation}\label{eq:HeatingRate}
\Gamma_{\rm h}=   \frac{q^2}{4m_I \hbar \omega_t} S_E(\omega=\omega_t),
\end{equation}
where $S_E(\omega)$ is  the fluctuation spectrum of the electric field at the position of the ion. Since $\Gamma_{\rm h}$ limits the time for performing coherent manipulations of the ion, a detailed understanding of $S_E(\omega)$,  its distance, frequency and temperature dependence, is of central importance for a further optimization and miniaturization of ion micro-traps.

As the trapping distance $d$ is decreased the ion becomes increasingly more sensitive to electric noise emerging from microscopic processes on the surface.
In~\cite{Safavi-Naini2011} we developed a microscopic model to describe the electric field noise, which is generated from a random distribution of adatoms on a gold surface.
In this case the field fluctuation spectrum for a planar trap geometry is given by 
\begin{equation}\label{eq:SESmu}
S_E(\omega_t)= \frac{3\pi}{4} \frac{\sigma}{(4\pi \epsilon_0)^2} \frac{ S_\mu(\omega_t)}{d^4}, 
\end{equation} 
where $\sigma$ is the surface density of dipoles and  $S_\mu(\omega)=\int_{-\infty}^\infty d\tau \langle \mu_z (\tau) \mu_z(0)\rangle  e^{i\omega \tau} $ is the spectrum of an individual fluctuating dipole. Eq.~\eqref{eq:SESmu} predicts the expected $d^{-4}$ scaling and together with Eq.~\eqref{eq:HeatingRate} it relates the ion heating rate to the microscopic dynamics of individual surface impurities. 

\begin{figure}[htb]
\includegraphics[width=0.48\textwidth]{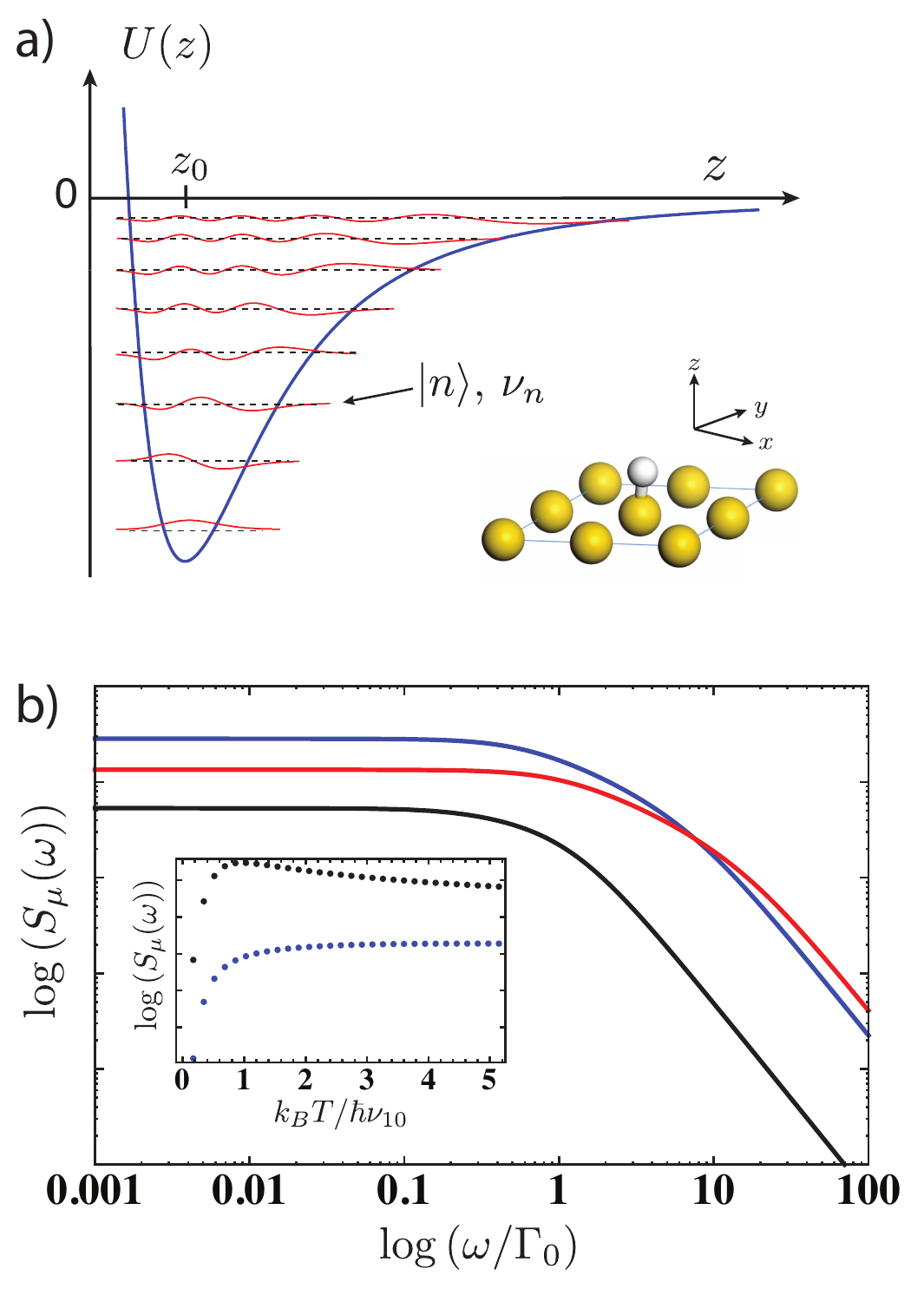}
\caption{\label{fig:FluctuatingDipole} (a) A schematic binding potential $U(z)$ of an adsorbate  on a bare Au surface as a function of the adatom-surface distance  $z$. The dotted lines are the bound states of the potential and the corresponding wavefunction is shown using solid lines. (b) The typical dependence of the phonon-induced dipole fluctuation spectrum of the adatom as a function of $\omega/\Gamma_0$, where $\Gamma_0$ is the characteristic transition rate from the first excited state to the ground state. The temperature is given in units of $\nu_{10}$, the separation between the ground and first excited vibrational states. See text and Ref.~\cite{Safavi-Naini2011} for more details.}
\end{figure}

\subsection{Phonon induced dipole fluctuations of adatoms}
Fig.~\ref{fig:FluctuatingDipole}~(a) shows a typical adatom-surface potential $U(z)$, which is attractive at large distances $z$ and has a sharp repulsive wall when the electronic wavefunctions of the adatom and the surface atoms start to overlap. The adatom-surface interaction is associated with a distortion of the electronic wavefunctions which results in an induced dipole moment  $\mu_z(z)$ perpendicular to the surface. At large distances one expects $\mu(z) ~\sim 1/z^4$~\cite{Dipole} and $\mu(z\approx z_0)$ can reach several Debye when the adatom touches the surface.

The potential $U(z)$ usually supports several bound vibrational states $|n\rangle$ with vibrational frequencies $\nu_n$ and the adatom can undergo phonon induced transitions between those vibrational states.  For $n>m$ the corresponding transitions rates are approximately given by \cite{Safavi-Naini2011}
\begin{eqnarray}
\label{eq:rates1}Ê
\Gamma_{n\to m}&=&\frac{\pi g(\nu_{nm}) }{3 \hbar M \nu_{nm}} \vert \langle n \vert  U^\prime(z)\vert m \rangle \vert ^2 \left(n(\nu_{nm})+1\right),\\
\label{eq:rates2}
\Gamma_{m\to n}&=&\frac{\pi g(\nu_{nm}) }{3 \hbar M \nu_{nm} }   \vert \langle n \vert  U^\prime(z)\vert m \rangle \vert ^2 n(\nu_{nm}).
\end{eqnarray}
Here $M$ is the surface atom mass, $g(\omega)$ is the phonon density of states (PDOS) and $n(\omega)=1/(e^{\hbar \omega/(k_BT)}-1)$ the thermal phonon occupation number, which are both evaluated at the vibrational transition frequency $\nu_{nm}=\nu_n-\nu_m>0$. Due to the  different average dipole moment 
$\mu_n=\langle n|\mu(z)|n\rangle$ associated with each vibrational state, absorption and emission of phonons creates a fluctuating dipole moment $\mu(t)$ as the adatom jumps between different levels $|n\rangle$.

\subsection{Dipole fluctuation spectrum}     
From the above considerations and a detailed knowledge of the adatom-surface potential $U(z)$, the induced dipole moment $\mu(z)$ and the phonon density of states $g(\omega)$, we can evaluate the dipole fluctuation spectrum and thereby the corresponding ion heating rate $\Gamma_{\rm h}$. The fluctuation spectrum is given by
\begin{equation}
S_\mu(\omega)=\int_{-\infty}^\infty d\tau \left( \langle \mu_z (\tau) \mu_z(0)\rangle -\langle \mu_z(0)\rangle^2\right) e^{i\omega \tau},
\end{equation}
where  $\mu_z=  \sum_n   \mu_{z,n} p_n$ and $p_n= |n\rangle\langle n| $ is the projection operator on the vibrational level $|n\rangle$. 

The typical dependence of $S_\mu(\omega)$ on frequency and temperature is shown in Fig.~\ref{fig:FluctuatingDipole}(b) establishing the rate $\Gamma_0\equiv\Gamma_{1\rightarrow0} (T=0)$ and the frequency $\nu_{10}\equiv (E_1-E_0)/\hbar$ as the relevant scales in the problem.  In~\cite{Safavi-Naini2011} we used approximate analytic model to estimate the relevant scales for a broad range of adatom species, but assuming a clean gold surface. 
By using a harmonic approximation for $U(z)$ and assuming that $g(\omega) \sim \omega^2$ we obtain~\cite{Safavi-Naini2011}  
\be
\label{eq:gamma}\nu_{10}\approx  \zeta \sqrt{\frac{U_0}{m z_0^2}},\qquad   \Gamma_0\approx \frac{1}{4\pi}\frac{\nu_{10}^4m}{v^3 \rho},
\ee  
where $\zeta\sim \mathcal{O}(1)$ is a numerical constant,  $m$ is the mass of the adsorbate, $\rho$ is the density of the slab, $U_0$ is the potential depth and $v$ is the speed of sound. 

In the following we consider a more realistic scenario and evaluate the potential modifications of the dipole fluctuation spectrum due to the presence of an additional ML of atoms on top of the Au surface. To do so we present in the following section exact numerical calculations adatom-surface potentials for the case of He and N monolayers, which provides us with an estimate for the minimal and maximal expected modification of the surface potential.   
%
Approximate analytic expressions for $\Gamma_0$ given in Eq.~\eqref{eq:gamma} allow us to extend these predictions for various monolayer-adatom combinations.

\section{Atom-monolayer-gold surface interaction} \label{methods} 
The asymptotic potential for a polarizable atom with dynamic polarizability $\alpha(\omega)$, which approaches a surface of dielectric constant $\epsilon$ is $U(z \gg z_0)\simeq -\frac{(\epsilon -1)}{(\epsilon + 1)} \frac{C_3}{z^3}$, where $C_3=\frac{1}{4\pi}\int{\alpha(i\omega)d\omega}$ and $z$ is the normal to the surface. As the atom approaches the surface, the interaction energy increases due to exchange of energy with phonons, leading to adsorption at some equilibrium distance $z_0$ near the surface, beyond which, for shorter distances of approach, the energy cost of electronic exchange between the electrons of the adsorbate atom and the bulk atom becomes too great to overcome, leading to a repulsive wall in the interaction potential. Below, we present {\it ab initio} calculations of the interaction potential energies normal to the substrate surface of H atoms with a ML of adsorbate atoms He and N on top of  the Au(111) surface.

\subsection{ML surface interaction potentials}
Total-energy calculations of bulk Au and Au(111) surfaces, with and without 
He and N adsorbate atoms, were performed using the spin-polarized density 
functional theory as implemented in the Vienna Ab initio Software Package 
(VASP) \cite{kresse1996}. The exchange correlation energy was calculated 
using the local gradient approximation (LDA) with the parametrization 
of Perdew and Wang (PWC) \cite{perdew1992}.

The interaction between valence electrons and ionic cores was described 
by the Projector Augmented Wave (PAW) method \cite{blochl1994,kresse1999}. 
The Au $5d^{10}6s^1$, N $2s^22p^3$ and He $1s^2$ electrons were treated 
explicitly as valence electrons in the Kohn-Sham (KS) equations and the 
remaining cores were represented by PAW pseudopotentials. The KS 
equations were solved using the blocked Davidson iterative matrix 
diagonalization scheme followed by the residual vector minimization 
method. The plane-wave cutoff energy for the electronic wavefunctions 
was set to 500~eV.

All structures were optimized with periodic boundary conditions
applied using the conjugate gradient method, accelerated using 
the Methfessel-Paxton Fermi-level smearing \cite{methfessel1989} 
with a width of 0.2~eV. The total energy of the system and 
Hellmann-Feynman forces acting on atoms were calculated with 
convergence tolerances set to $10^{-3}$~eV and 0.01~eV/{\AA}, 
respectively. Structural optimizations and properties calculations 
were carried out using the Monkhorst-Pack special $k$-point 
scheme \cite{monkhorst1976} with $11\times11\times11$ and 
$7\times7\times1$ meshes for integrations in the Brillouin 
zone (BZ) of bulk and slab systems, respectively. 

The supercell consisted of a three-layer thick gold slab with 
$(111)$ orientation and a $p(2\times2)$ mesh unit, covered by 
He or N adsorbate atoms on one side of the slab model. 
The calculated lattice constant of bulk Au was 4.06 \AA, in close 
agreement with the experimental value of 4.0780~\AA~at $25^{\circ}$C 
\cite{dutta1963}. The lattice parameters of the $p(2\times2)$ Au(111) 
surface constructed by cleaving the optimized bulk structure were 
$a=b=5.74$~\AA~and $c=25.00$~\AA, with {\it ca.} $20.00$~\AA~vacuum 
separating slabs, and $\alpha=\beta=90^{\circ}$ and 
$\gamma=120^{\circ}$. 
Although a large vacuum region ({\it ca.} 20~\AA) was used between 
periodic slabs, the creation of dipoles upon adsorption of atoms on 
only one side of the slab can lead to spurious interactions 
between the dipoles of successive slabs. In order to circumvent 
this problem, a dipole correction was applied by means of a dipole 
layer placed in the vacuum region following the method outlined 
by Neugebauer and Scheffler \cite{neugebauer1992}. 

The ${\rm He}/{\rm Au}(111)$ and ${\rm N}/{\rm Au}(111)$ interaction 
potentials were calculated by gradually moving a single He or N atom 
along the $z$-axis normal to the Au(111) surface.

\subsection{Atomic adsorption on Au(111)}

Four different atomic adsorption sites are possible onto a 
Au(111) surface: 1) a bridge site between two gold atoms, 2) on
top of a gold atom, 3) in a hollow site between three gold atoms, 
termed an hexagonal close packed (hcp) site when there is a gold 
atom in the layer directly beneath the surface layer, or 4) termed 
a face-centered cubic (fcc) site when there is a hole in the layer 
directly beneath the surface layer.   

Total-energy calculations indicate that a single He atom adsorbs 
preferentially at the bridge site ($E=-48.760$~eV), slightly more 
energetically favorable than at the top site ($-48.756$~eV), the 
fcc site ($-48.755$~eV) and the hcp site ($-48.749$~eV). The 
elongated equilibrium He--Au bond distance of 3.58~\AA~suggests 
that He at the bridge site is weakly physisorbed to the Au(111) 
surface. For the adsorption of a single N atom, the fcc site is 
energetically preferred ($E=-70.393$~eV) over the hcp site 
($E=-69.792$~eV), the bridge site ($-69.098$~eV) and the 
top site ($E=-66.510$~eV). Contrasting with the He adsorbate, 
the N atom occupying the fcc site appears chemisorbed to the 
Au(111) surface with a short N--Au bond distance of 2.05~\AA. 

\begin{figure}[htb]
\begin{center}
\includegraphics[width=0.5\textwidth]{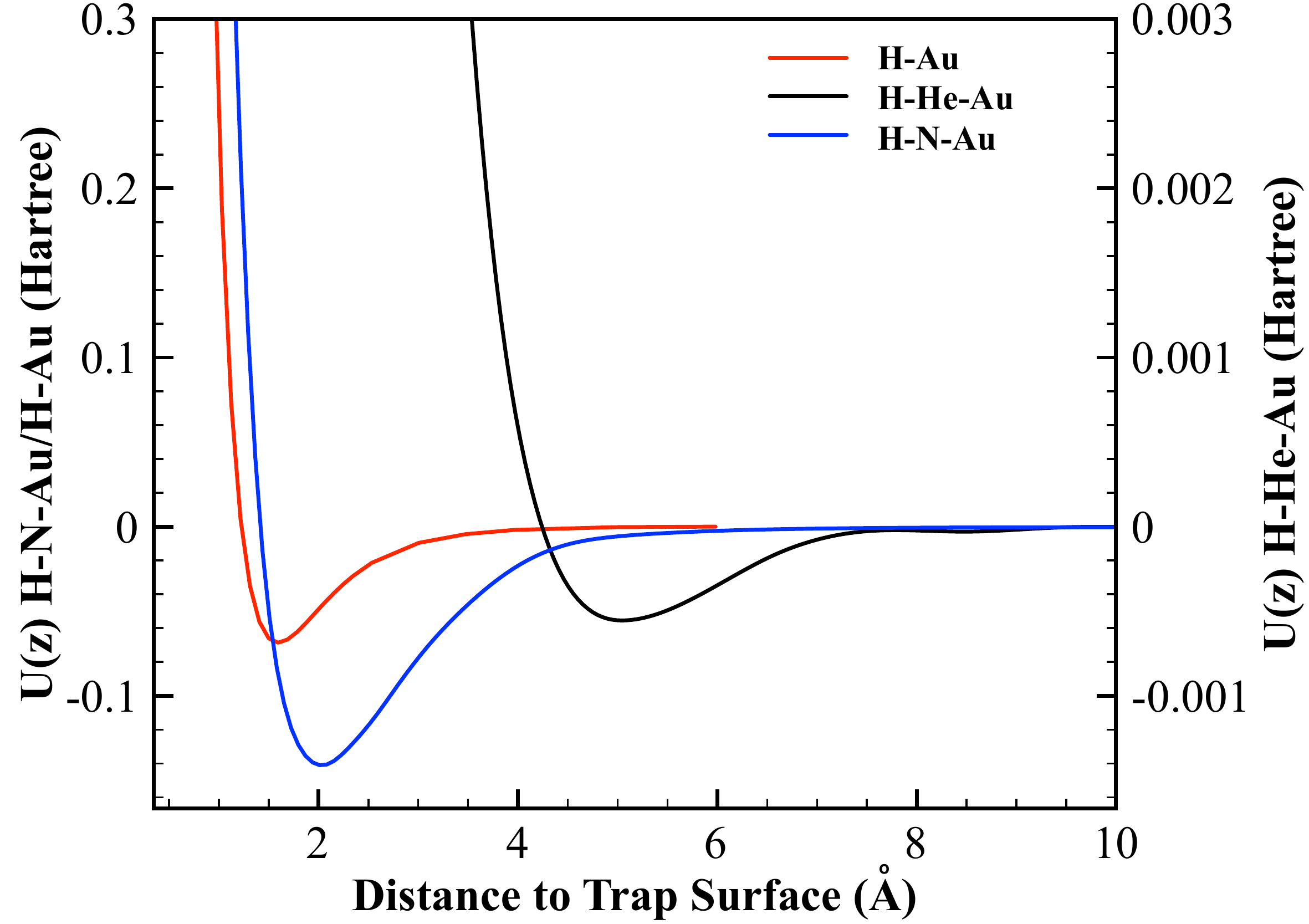}
\end{center}
\caption{\label{fig:mlpot} The binding potentials for H adsorbate 
atoms on bare Au surface (dotted) and Au surface covered with
He (solid) or N (dashed) monolayers.  The peak of the local potential at the position of the first Au layer, in direct contact with the N(He) ML,  does (not) vary appreciably indicating the formation of a sizable (negligible) dipole moment.
}
\end{figure}

Fig.~\ref{fig:mlpot} shows the adsorbate potentials for the two monolayers as well as the bare Au surface.  Note that we have shifted the potentials so that $E(z_0\rightarrow \infty)=0$. The presence of the weakly adsorbed He results in a much shallower potential while the chemisorbed N has the opposite effect, creating a deeper well that support more bound states. 
\subsection{Atomic diffusion on Au(111)}

Although our primary focus in this work is on dipole fluctuations induced by atomic motion perpendicular to the surface, the numerical data obtained in the previous part allows us with no additional effort to evaluate the diffusion rates of adatoms parallel to the surface.  Such data could be relevant to other, diffusion related noise processes~\cite{Diffusion1,Diffusion2,Diffusion3}  under identical conditions.  

The surface diffusion coefficient is given by
$D=\frac{\sqrt 3}{4} r_0^2 \Gamma$ for an fcc(111)
surface, where $r_0$ is the lattice parameter [2.87~\AA~for Au(111)]
and $\Gamma$ is the jump rate.
Two different diffusion regimes exist, namely a thermally
activated regime and a quantum tunneling regime.

In the thermally activated regime, the thermal jump rate, $\Gamma_{\rm therm}$, 
can be calculated through the Arrhenius formula \cite{arrhenius1889}, 
\begin{equation}
\Gamma_{\rm therm} = \gamma \exp \left ( -\frac{E_a}{k_{\rm B}T} \right ),
\end{equation}
where $E_a$ is the activation energy barrier, $k_{\rm B}$ is the Boltzmann 
constant, $T$ the temperature, and $\gamma$ is a prefactor which 
contains dynamical quantities; $\gamma \approx 10^{12}-10^{13}$~Hz for 
most surfaces \cite{ovesson2001}. The diffusion of a He atom on Au(111) 
is expected to be nearly barrierless due to their weak physisorption 
interaction.

For the diffusion of a single N atom between adjacent stable fcc sites 
on Au(111) the activation energy was calculated in this study to be 
$E_a=0.17$~eV using the nudge elastic band method within the DFT 
framework. This result is close to the value of $\approx 0.10$~eV 
determined previously from experiment and theory for the diffusion of 
Cr on Au(111) surface \cite{ohresser2005}. 
Following a simple Redhead's analysis of the migration temperature 
on solid surfaces \cite{masel1996}, we estimate the temperature 
necessary for a N adatom to overcome this energy barrier to be 
$\approx 65$~K~[$E=0.06~T$~kcal mol$^{-1}$ K$^{-1}$]. Let us note that Redhead's
law predicts the migration temperature of Cr adatoms on Au(111) to 
be 39~K, in excellent agreement with experimental 
findings \cite{ohresser2005}. 
The thermally-activated diffusion jump rate of N on Au(111) at 65 K 
is calculated to be $\Gamma_{\rm therm}=0.3$~Hz for an activation 
energy of 0.17~eV~and a prefactor $\gamma=5 \times 10^{12}$~Hz; this 
corresponds to a surface diffusion coefficient 
$D=1.2 \times 10^{-16}~{\rm cm}^2{\rm s}^{-1}$. In the temperature range 
$60-70$~K, $\Gamma_{\rm therm}$ and $D$ vary from $2.6 \times 10^{-2}$ to 
2.9~Hz and from $9.3 \times 10^{-18}$ to 
$1.0 \times 10^{-15}$ ~${\rm cm}^2{\rm s}^{-1}$, respectively. 

Below $\approx 65$~K, classical thermally activated surface 
diffusion can be excluded as the origin of the N adatom 
diffusion and the site-to-site hopping rate is controlled 
by quantum tunneling. In this temperature regime the de Broglie
wavelength of a N adatom of mass $m$, i.e. 
$\lambda_{\rm B}=2 \pi \hbar / \sqrt{3mk_{\rm B}T}$, is comparable 
to the distance separating adjacent adsorption sites on the Au(111) 
surface, therefore making quantum tunneling possible. 
For the fundamental energy level, the tunneling jump rate, 
$\Gamma_{\rm tunnel}$, can be derived for a one-dimensional parabolic 
double-well potential and approximated by \cite{merzbacher1970}
\begin{equation}
\Gamma_{\rm tunnel} = \frac{2 \omega}{\pi^{3/2}} 
 \sqrt{\frac{2E_a}{\hbar \omega}} \exp 
 \left ( -\frac{2 E_a}{\hbar \omega} \right ),
\end{equation}
where $\omega=\sqrt{2E_a/mb^2}$ and $b$ is the barrier width 
[$b=0.8$~\AA~for Au(111)]. Therefore, the activation energy $E_a=0.17$~eV
corresponds to $\Gamma_{\rm tunnel}=66.9$~Hz. If the calculated 
$E_a$ is considered to be accurate within $\pm 10$~meV, 
$\Gamma_{\rm tunnel}$ can vary in the range $\approx 31.9-143.3$~Hz.

\subsection{Work functions and surface dipoles}

\begin{figure}[htb]
\begin{center}
\includegraphics[width=0.5\textwidth]{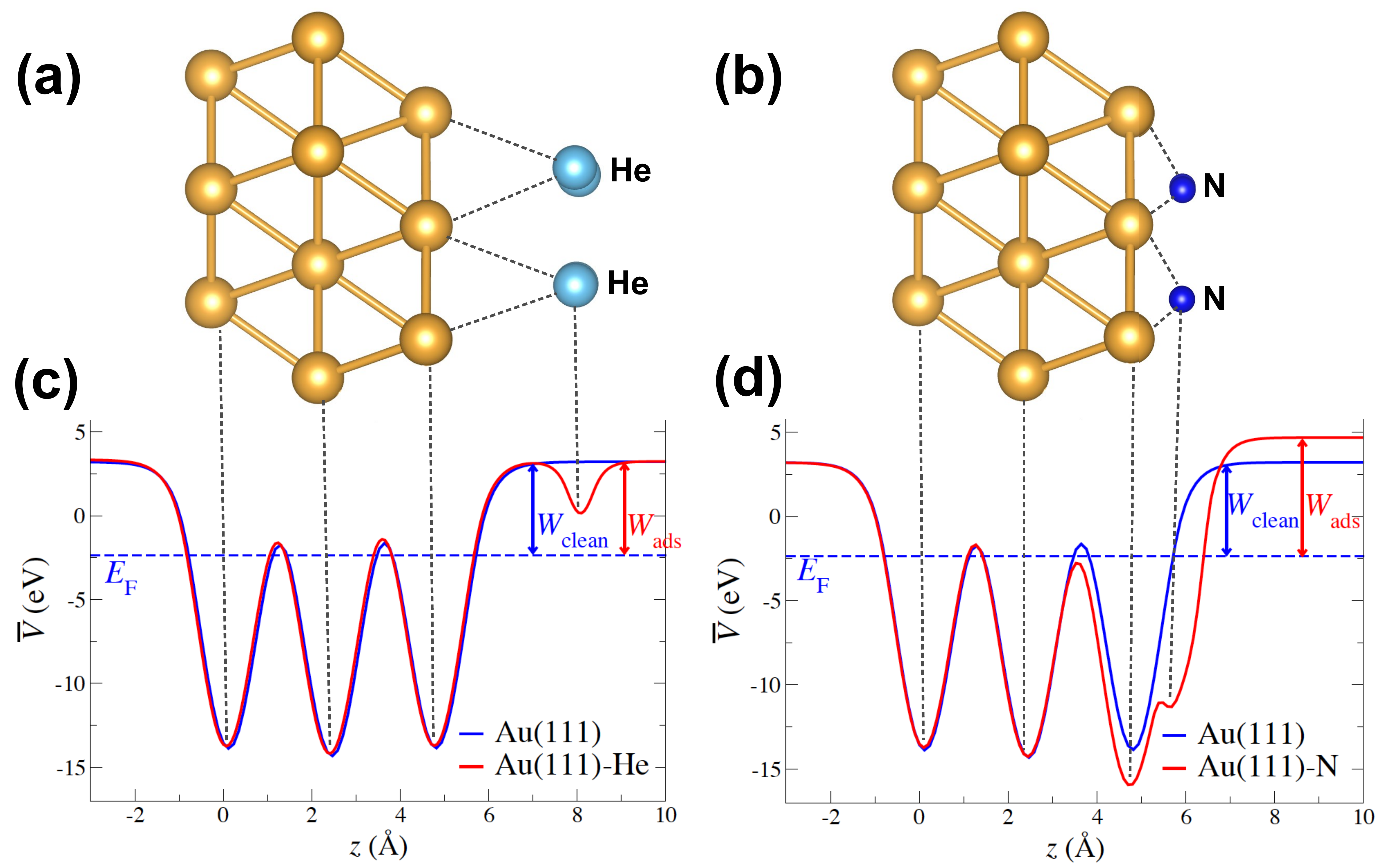}
\end{center}
\caption{\label{fig1} Au(111) slab models covered by a one-sided 
monolayer (1~ML) of (a) He atoms and (b) N atoms. The variation of 
the corresponding plane averaged electrostatic potential 
$\overline{V}(z)$ along the $z$-axis normal to the surface is 
represented for (c) He-covered and (d) N-covered Au(111) slabs, 
together with a clean Au(111) slab reference. The Fermi energy 
$E_{\rm F}$ (horizontal dashed line) and the work functions (vertical 
arrows) of the clean and adsorbate-covered slabs, $W_{\rm clean}$ 
and $W_{\rm ads}$, respectively, are also represented.  
}
\end{figure}

The work function, $W$, is defined as the minimum energy required 
to remove an electron from a solid to the vacuum region in the vicinity 
of the solid surface and is given by:
\begin{equation}
 W = \overline{V}(\infty)-E_{\rm F},
\end{equation}
where $\overline{V}(\infty)$ is the plane-averaged electrostatic 
potential in the vacuum at a distance where the microscopic potential 
has reached its asymptotic value and $E_{\rm F}$ is the Fermi energy.

The electrostatic potential $V(x,y,z)$ on a grid in real space can 
be obtained from a self-consistent electronic structure calculation 
using a plane wave basis set. Assuming that the surface normal is 
oriented along the $z$-axis, one can define a plane averaged potential
\begin{equation}
\overline{V}(z) = \frac{1}{\cal A} \iint_{\rm cell}V(x,y,z)dxdy,
\end{equation}
where $\cal A$ is the supercell surface area. The asymptotic value 
$\overline{V}(\infty)$ can be extracted by plotting the variation of 
$\overline{V}$ as a function of $z$, as shown in Fig.~\ref{fig1} 
for a clean Au(111) surface and for a Au(111) surface covered 
by 1~ML of He and N atoms. 

The calculated Fermi energy and electrostatic potential in the vacuum 
for the clean Au(111) surface are $E_{\rm F}=-2.36~{\rm eV}$ and 
$\overline{V}(\infty)=+3.21~{\rm eV}$. This corresponds to a work function  
$W_{\rm clean}=5.57~{\rm eV}$, in good agreement with the experimental 
value of 5.50~eV recently measured by Br\"oker {\it et al.} for this 
Miller index plane \cite{broker2008}.

Adsorption of one monolayer of He atoms at bridge sites and N atoms 
at fcc sites onto the Au(111) surface results in 
$\overline{V}(\infty)=+3.24~{\rm eV}$ and 
$\overline{V}(\infty)=+4.68~{\rm eV}$, respectively. Thus, the
work functions of He- and N-covered Au(111) surfaces are 
$W_{\rm ads}=5.60$ and $7.04~{\rm eV}$. 

In order to analyze the change of the work function upon atomic 
adsorption, we define the variation $\Delta W = W_{\rm ads} - W_{\rm clean}$.
For the weakly physisorbed He monolayer this variation is negligible 
$(\Delta W=0.03~{\rm eV})$, unlike in the case of the chemisorbed N 
monolayer $(\Delta W=1.47~{\rm eV})$. The variation of the work 
function results from the change in the surface electric dipole 
caused by adsorption of the adatoms.
Simple electrostatics gives the relation \cite{jackson1975}
\begin{equation}
\Delta W = \frac{e \Delta \mu}{\epsilon_0 A},
\end{equation}
where $A$ is the surface area taken up by one adatom, $\epsilon_0$ 
is the electric permittivity of free space and $\Delta \mu$ 
is the change in surface dipole that occurs upon atomic adsorption, 
normalized per adatom. $\Delta \mu$ corresponds to the $z$-component 
of the dipole moment directed along the surface normal, since
only this component affects the work function. Since four adatoms 
form 1~ML covering the supercell surface area, the surface area by 
adatom can be approximated by $A\approx{\cal A}/4$. We can now estimate the
induced dipole moment for the case of He and N adsorbates. Using a DFT 
unit cell area of
${\cal A}\approx 41$~\AA$^2$ we find $\mu_{\rm He}\approx 0.03$~D 
while $\mu_{\rm N}\approx1.60$~D.

It should be noted that the major contribution to the surface dipole 
results from the charge reordering associated with the formation 
of the chemical bonds between the metal surface and the adatoms. 
This contribution is foremost determined by the nature of the 
chemical bonds, but can also be modified by the packing density 
of the adatoms.

\subsection{Phonon Density of States in Presence of the Monolayer Adsorbates}

\begin{figure}[htb]
\begin{center}
\includegraphics[width=0.5\textwidth]{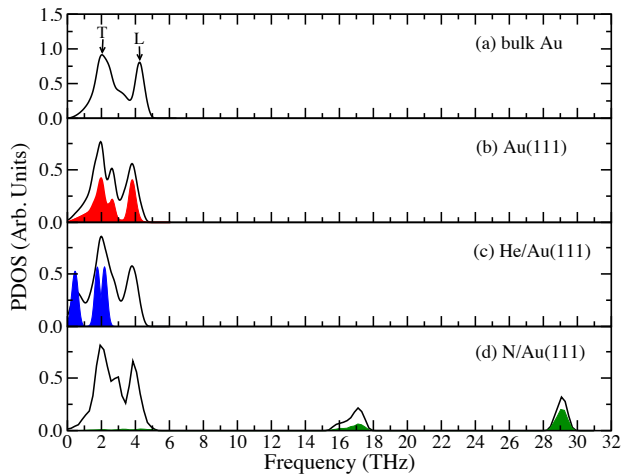}
\end{center}
\caption{ \label{fig:PDOS} Phonon density of states (PDOS): (a) bulk 
Au, (b) bare Au$(111)$ surface, (c) 1ML He-covered Au(111) surface, and 
(d) 1ML N-covered Au(111) surface. The curves in black are the calculated 
total PDOS and the shaded areas in red, blue, and green represent the 
partial PDOS projected to the surface atoms, He atoms, and N atoms, 
respectively.} 
\end{figure}

Phonon density of states (PDOS) were calculated by solving the dynamical 
matrix for bulk Au, clean Au(111) surface, and adsorbates (e.g., He or N) 
on the Au(111) surface as shown in Fig.~\ref{fig:PDOS}. A 
$(2 \times 2 \times 2)$ supercell was adopted to obtain the force constant 
matrix of bulk Au that can be  derived from the Hellmann-Feynman forces 
obtained from the DFT calculations using VASP \cite{kresse1996}. 
The calculated PDOS of the bulk Au shows two main peaks represented 
by ``T" and ``L" that are in good agreement with previous experimental 
results \cite{lynn}. According to Lynn {\it et al.} \cite{lynn}, the 
longitudinal (L) and transverse (T) phonon modes of bulk Au are 4.61 THz 
and 2.75 THz, respectively, at a high symmetry point (X) in the 
Brillouin zone.   

The calculated surface PDOS using the $(2 \times 2)$ supercells are 
depicted in Fig.~\ref{fig:PDOS}(b) for a clean Au(111) surface as well 
as for 1ML of adsorbates (He or  N) on the Au(111) surface.  There are 
three prominent peaks in the calculated PDOS of the clean Au(111) surface, 
mainly contributed by the partial PDOS projected onto the surface atoms 
(red shaded area). The previous experimental study has identified four 
surface modes (2.31 THz, 3.5 THz, 4.0 THz, and 4.3 THz) at a high 
symmetry point (K) in the surface Brillouin zone \cite{ponjee}.     

The calculated PDOS of the 1ML He-covered Au(111) surface suggests  
a very weak interaction between He and the metal surface, providing no
evidence of stretching or wagging modes of He atoms with respect to the 
metal surface. However, the partial PDOS projected onto the He atoms 
(blue shaded area) reveals the possible lattice modes of the 1ML He atoms 
physisorbed on the surface as shown in Fig.~\ref{fig:PDOS}(c).

Contrasting with the 1ML He-covered Au(111) surface, two additional 
peaks appear at high frequency above 5 THz for the 1ML N-covered Au(111) 
surface due to the wagging and stretching modes of N atoms attributed 
to the strong interaction with the metal surface, as shown in 
Fig.~\ref{fig:PDOS}(d).  
The green shaded area represents the partial PDOS projected onto the N atoms.

\section{Noise spectrum with ML present}\label{sec:noise} 
Let us now study the impact of the two different types of MLs described above on the dipole fluctuation spectrum of adatoms as discussed in Sec.~\ref{sec:Heating}.  
We first note that due to the high reactivity and low mass of H adsorbates -- chosen in the previous section to reduce the DFT computational cost -- the depth of the binding potential $U_{0, bare}=0.068$~Hartree and vibrational frequencies $\nu_{10,bare}/2\pi\approx 40$~THz are high. Therefore, for H adatoms thermally activated processes at room temperature do not play a significant role. Instead we use the potentials shown in Fig.~\ref{fig:mlpot} together with the mass scaling relations in   Eq.~\eqref{eq:gamma} to evaluate the noise spectrum for a more realistic set of adsorbates with masses around $m\sim 100$ amu.

\begin{figure}[h]
\begin{center}
\includegraphics[width=0.49\textwidth]{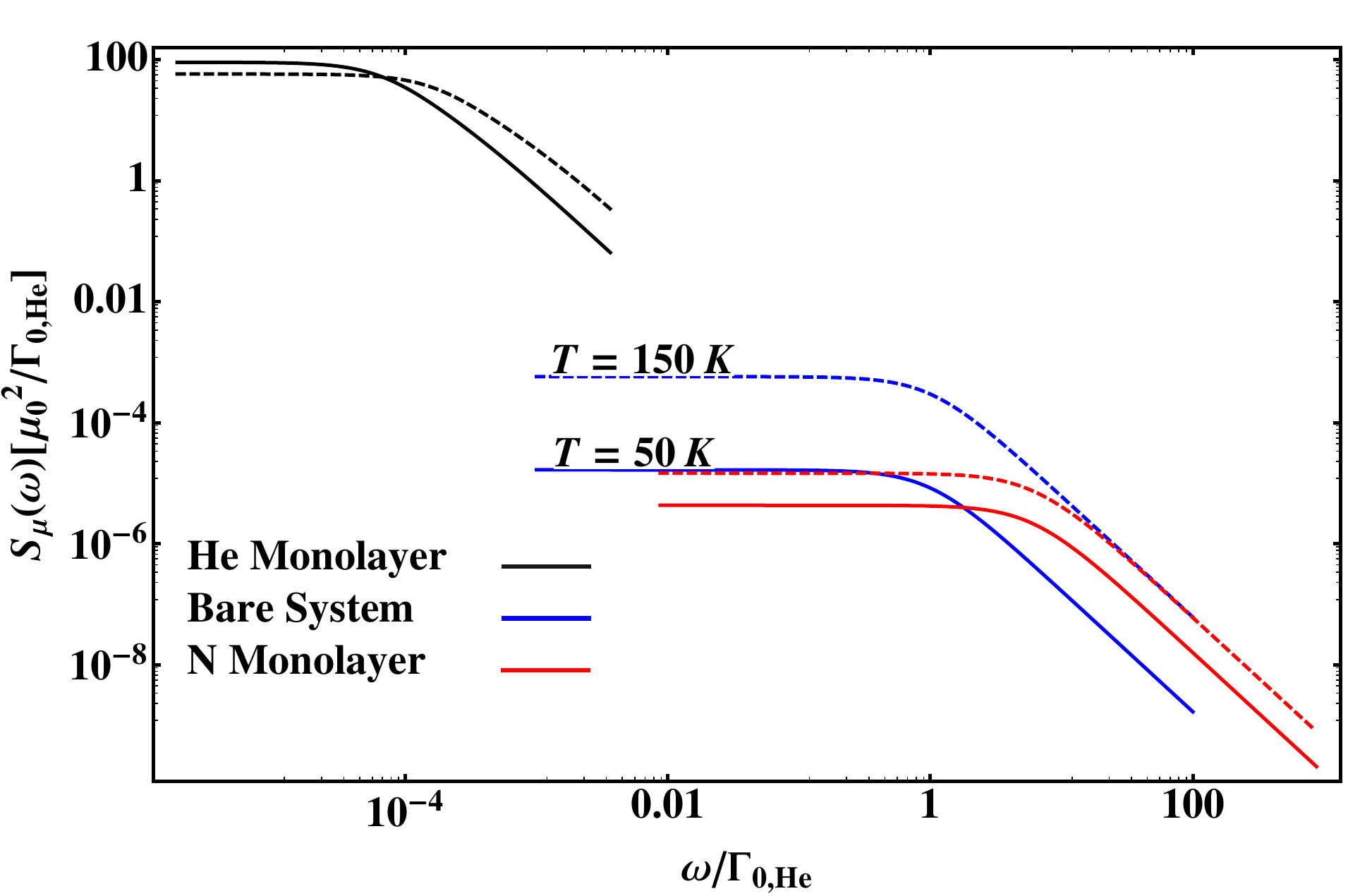}
\end{center}
\caption{ \label{fig:spec} The dipole fluctuation spectrum $S_\mu (\omega)$ as a function of $\omega/\Gamma_{0, Au}$ for He monolayer (black), N monolayer (red) and bare system (blue). For each system the solid curve corresponds to $T=50$~K while the dashed curve corresponds to $T=150$~K. The frequency $\omega$ is scaled by $\Gamma_{0, Au}$, the characteristic rate corresponding to He monolayer coverage. The spectrum is given in units of $\mu_0^2/\Gamma_{0, Au}$, where $\mu_0$ is the induced dipole moment for the ground state for each system.  
}
\end{figure}

In Fig.~\ref{fig:spec}  we plot the dipole fluctuation spectrum $S_\mu(\omega)$ for a bare Au surface and in the presence of a He and N monolayer. For the results shown in Fig.~\ref{fig:spec} we use the potentials shown in Fig.~\ref{fig:mlpot} with an adatom mass of 100 amu and two different temperatures of $T=50$ K (solid) and $T=150$ K (dashed). For the bare Au surface where $z_0\approx1.59$~\AA \ these values correspond to $\nu_{10}/2\pi \approx 4.5$~THz, $\Gamma_{0,Au}/2\pi\approx 2.2$~THz and ratios $k_BT/\hbar \nu_{10} \approx 0.20$ and 0.70 respectively.  For a qualitative discussion of the strong modifications of the spectrum  in the presence of a ML, we consider low temperatures, where the adatom potential can be approximated by only two vibrational level system and  
\be \label{eq:SpectrumTLS}
S_\mu(\omega)=(\mu_1-\mu_0)^2\frac{2 \Gamma_0}{\Gamma_0^2+\omega^2}e^{-\hbar \nu_{10}/k_BT }.
\ee 
In Sec.~\ref{methods}, we have found that due to its low reactivity, the He monolayer results in a significantly shallower well depth, $U_{0, He}=0.00055$~Hartree, shifting the minimum to $z_{0, He}\approx 5$~\AA. Both effects lower the vibrational frequency, $\nu_{10, He}\approx 0.4$~THz, and lead to a drastic reduction of the characteristic phonon transition rate $\Gamma_{0, He}/2\pi \approx 140$~MHz. From Eq.~\eqref{eq:SpectrumTLS} this results in an increase of the low frequency noise level, but reduces the noise at frequencies $\omega \gg \Gamma_0$. 

The more reactive N monolayer results in a deeper potential well with $U_{0, N}=0.141$~Hartree, while only slightly affecting the equilibrium distance, $z_{0, N}\approx 2$~\AA. This leads to exactly the opposite effect, increasing the vibrational frequency, $\nu_{10, N}\approx 5.3$~THz as well as $\Gamma_{0,N}/2\pi\approx 3.9$~THz. As also seen in Fig.~\ref{fig:spec}, the stronger binding therefore reduces the noise in the low frequency regime. Additionally the low frequency regime of the spectrum extends over a much larger frequency range.

Finally, let us briefly comment on the modified PDOS. Our estimates of $\Gamma_0$ so far have been based on Eq.~\eqref{eq:gamma}, which assumes $g(\omega)\propto \omega^2$.  This scaling  ignores the detailed structure of the PDOS at high frequencies and overestimates $\Gamma_0$ for $\nu_{10}$ above $\sim 1$~THz. Although the variations in PDOS shown in Fig.~\ref{fig:PDOS} are in general less relevant and only important for adatoms with high vibrational frequencies, it seems feasible to identify ML species, for which the PDOS is peaked at the vibrational frequencies $\nu_{10}$, thereby increasing $\Gamma_0$ and suppressing the low frequency noise level.

\section{Summary and Conclusions}\label{sec:conclusions}
In summary, we have calculated the noise induced heating in ion traps due to randomly-fluctuating adatom dipoles in the presence of a single ML of atomic species on Au(111) surface. Precise DFT calculations of surface potentials for physisorbed and chemisorbed ML species provided us with accurate data for surface potentials, from which the effects on the noise could be evaluated. We have found that within our noise model, the presence of surface contamination can lead to opposite effects of enhancing or reducing the noise level, depending on the reactivity of the ML species as well as the frequency range.

Although an exact quantitative comparison between experiments, for example, in Ref. \cite{Allcock} and \cite{wineland2011}, is beyond the scope of  this work, our current analysis points the direction for a more refined understanding of anomalous heating in ion traps, requiring the combined knowledge of atomic surface physics as well as the modeling of different noise processes.  In particular, the combined data on surface potentials, PDOS and diffusion rates obtained in this work could in the future serve as a common input to evaluate and compare different alternative noise models. This can lead to better understanding of the noise mechanism, allowing ion traps to be used as exquisite probes of surface reactivity.

\section{Acknowledgments}
Sandia National Laboratories is a multi-program laboratory managed and operated by Sandia Corporation, a wholly owned subsidiary of Lockheed Martin Corporation, for the U.S. Department of Energy\textquoteright s National Nuclear Security Administration under contract DE-AC04-94AL85000.
P.R. acknowledges support by the Austrian Academy of Sciences, and the Austrian Science
Fund (FWF) through the START grant Y 591-N16.
This work was supported by NSF through a grant to ITAMP at the Harvard-Smithsonian Center for Astrophysics.



\end{document}